\documentclass[letterpaper,titlepage,12pt]{elsarticle}
\usepackage[colorlinks=true,urlcolor=blue,citecolor=magenta]{hyperref}
\usepackage{amssymb,amsmath,amsfonts}
\usepackage{epsfig}
\usepackage{epsfig}
\usepackage{graphicx}
\usepackage{epstopdf}
\usepackage{caption}
\usepackage{subcaption}
\usepackage{amscd}
\usepackage{amsthm}
\usepackage{latexsym}
\usepackage{amsbsy}
\usepackage{bbm}
\usepackage[english]{babel}
\usepackage{psfrag}
\usepackage{tabularx}

\def\be{\begin{equation}}
\def\ee{\end{equation}}
\def\bea{\begin{eqnarray}}
\def\eea{\end{eqnarray}}

\allowdisplaybreaks

\def\clock{{\count0=\time
           \divide\count0 60
           \ifnum\count0<10 0\fi\the\count0
           \multiply\count0 -60 \advance\count0 \time
           :\ifnum\count0<10 0\fi \the\count0
         }}
\newcommand{\timestamp}{{\small\vbox{\hbox{\tt\jobname.tex}
\hbox{\the\day/\the\month/\the\year, \clock}}}}

\newcommand{\spa}{\quad , \quad}

\def\sp{\;\;\;,\;\;\;}

\numberwithin{equation}{section}

\begin{document}

\begin{titlepage}
\rightline{\vbox{CCTP-2014-16, CCQCN-2014-39   \phantom{ghost} }}

 \vskip 1.8 cm

\centerline{\LARGE \bf
Lifshitz Space-Times for Schr\"odinger Holography}

\vskip 1.5cm

\centerline{\large {{\bf Jelle Hartong$^1$, Elias Kiritsis$^{2,3}$, Niels A. Obers$^1$}}}

\vskip 1.0cm

\begin{center}
\sl $^1$ The Niels Bohr Institute, Copenhagen University,\\
\sl  Blegdamsvej 17, DK-2100 Copenhagen \O , Denmark.\\
\sl $^2$ Crete Center for Theoretical Physics, Department of Physics, University of Crete,\\
 71003 Heraklion, Greece.\\
\sl $^3$ APC, Universit\'e Paris 7,
CNRS/IN2P3, CEA/IRFU, Obs. de Paris,
Sorbonne Paris\\ Cit\'e, B\^atiment Condorcet, F-75205, Paris Cedex 13, France,
(UMR du CNRS 7164).
\vskip 0.4cm
\end{center}
\vskip 0.6cm

\centerline{\small\tt  hartong@nbi.dk, kiritsis@physics.uoc.gr, obers@nbi.dk}

\vskip 1.3cm \centerline{\bf Abstract} \vskip 0.2cm \noindent

We show that asymptotically locally Lifshitz space-times are holographically dual to field theories that exhibit Schr\"odinger invariance. This involves a complete identification of the sources, which describe torsional Newton-Cartan geometry on the boundary and transform under the Schr\"odinger algebra. We furthermore identify the dual vevs from which we define and construct the boundary energy-momentum tensor and mass current and show that these obey Ward identities that are organized by the Schr\"odinger algebra. We also point out that even though the energy flux has scaling dimension larger than $z+2$, it can be expressed in terms of computable vev/source pairs.

\end{titlepage}

\newcommand{\red}[1]{{\color{red} #1 \color{black}}}
\newcommand{\blue}[1]{{\color{blue} #1 \color{black}}}
\newcommand{\green}[1]{{\color{green} #1 \color{black}}}
\newcommand{\yellow}[1]{{\color{yellow} #1 \color{black}}}

\newcommand{\EK}[1]{{\red{EK: #1}}} 
\newcommand{\JH}[1]{{\blue{JH: #1}}} 
\newcommand{\NO}[1]{{\green{NO: #1}}} 



\section{Introduction}

Many systems in nature exhibit critical points with non-relativistic scale invariance. Such systems typically have Lifshitz symmetries, which include anisotropic scaling between time and space, characterized by a dynamical critical exponent $z$. A larger symmetry group that also displays non-relativistic scale invariance, which contains the Lifshitz group, is the Schr\"odinger group which possesses as additional symmetries the Galilean boosts and a particle number symmetry. Over the last six years, following the success of holography in describing strongly coupled relativistic field theories, there has been a growing interest in applying similar techniques to strongly coupled systems with non-relativistic symmetries
\cite{Son:2008ye,Balasubramanian:2008dm,Kachru:2008yh,Taylor:2008tg}. In this letter we show that, when applying holography to asymptotically locally Lifshitz space-times,  the resulting dual field theories exhibit Schr\"odinger invariance.

Our development builds on the recent works \cite{Christensen:2013lma,Christensen:2013rfa} in which, for a specific action supporting $z=2$ Lifshitz geometries, the Lifshitz UV completion was identified by solving for the most general solution near the Lifshitz boundary. A key ingredient in these works is the use of a vielbein formalism enabling the identification of all the sources as the leading components of well-chosen bulk fields. This includes in particular two linear combinations of the timelike vielbein and the bulk gauge field, where one asymptotes to the boundary timelike vielbein and the other to the boundary gauge field. The latter plays a crucial role in the resulting geometry that is induced from the bulk onto the boundary, which in \cite{Christensen:2013lma,Christensen:2013rfa} was shown to be a novel extension of Newton--Cartan geometry with a specific torsion tensor, called torsional Newton--Cartan (TNC) geometry.  By considering the coupling of this geometry to the boundary field theory, the vevs dual to the sources were computed and moreover their Ward identities were written down in a TNC covariant form. Among others, this includes the gauge invariant boundary energy-momentum tensor, from which the energy density, momentum flux, energy flux and stress can be computed by appropriate tangent space projections. 

We consider in this work a large class of Lifshitz models for arbitrary values of $z$ (focussing on $1< z \leq 2$), where we find that the above results generalize, and moreover that there is an underlying Schr\"odinger symmetry that acts on the sources and vevs,  revealing that the boundary theory has a Schr\"odinger invariance. The arguments of this letter are furthermore supported by a complementary analysis of bulk versus boundary Killing symmetries in \cite{Hartong:2014pma}. This approach employs the TNC analogue of a conformal Killing vector, which was identified for the first time in \cite{Christensen:2013rfa} by deriving the conditions for the boundary theory to admit conserved currents. We also note that details of the present work and \cite{Hartong:2014pma} along with further results are given in \cite{Hartong:2015wxa,Hartong:2014}. Finally in a companion paper \cite{Bergshoeff:2014uea} it is shown how to obtain all the details of the TNC geometry by gauging the Schr\"odinger algebra. The notation among the papers \cite{Hartong:2014pma,Bergshoeff:2014uea,Hartong:2015wxa,Hartong:2014} together with the current one is fully compatible.

Our results are of relevance to the general  study of holography for Lifshitz space-times \cite{Ross:2009ar,Ross:2011gu,Baggio:2011cp,Mann:2011hg,Griffin:2011xs,Christensen:2013lma,Christensen:2013rfa,Chemissany:2014xpa,Chemissany:2014xsa}%
\footnote{See also \cite{Guica:2010sw,Hartong:2010ec,Guica:2011ia,Hartong:2013cba,Compere:2014bia,
Andrade:2014iia} for related recent work on Schr\"odinger and warped AdS$_3$ space-times.}%
, which is interesting in its own right as a tractable example of  non-AdS space-times for which it is possible to construct explicit holographic techniques. But another more concrete motivation is, as remarked above, the application of these ideas and results to condensed matter type systems. In this connection, we note that TNC geometry has recently appeared in relation to field theory analyses of problems with strongly correlated  electrons, such as the quantum Hall effect (see e.g. \cite{Gromov:2014vla,Geracie:2014nka,Brauner:2014jaa,Geracie:2014zha} following the earlier work \cite{Son:2013rqa} that introduced NC geometry to this problem). In parallel to the renewed development of relativistic fluid and superfluid dynamics that was initiated and inspired by the fluid/gravity correspondence \cite{Policastro:2001yc,Bhattacharyya:2008jc}, we expect that our holographic approach to Lifshitz space-times will lead to further novel insights into the dynamics and hydrodynamics of non-relativistic field theories.

{\bf Note added:}
While this letter was being finalized, the preprint \cite{Jensen:2014aia} appeared on the arXiv, which appears to have some overlap with our results regarding coupling to TNC backgrounds.

\section{EPD model and asymptotically locally Lifshitz solutions \label{EPD} \label{sec:alif} }

We consider a holographic theory with a metric $g_{MN}$, a massive vector field $B_M$ and a scalar $\Phi$ (Einstein-Proca-Dilaton (EPD) theory) with the following bulk action%
\footnote{We use capital roman indices $M=(r,\mu)$ for the four-dimensional bulk space-time, with boundary space-time indices
$\mu$  and spatial tangent space indices $a=1,2$.} 
\begin{equation}\label{eq:action}
 S=\int d^4x\sqrt{-g}\left(R-\frac{1}{4}Z(\Phi)F^2-\frac{1}{2}W(\Phi)B^2-\frac{1}{2}(\partial\Phi)^2-V(\Phi)\right)\,,
\end{equation}
where $F=dB$. 
The Lagrangian has a broken $U(1)$ gauge symmetry signaled by the mass term of $B_M$.
The functions $Z(\Phi)$ and $W(\Phi)$ are positive but otherwise arbitrary functions of the scalar field $\Phi$ and
the potential $V(\Phi)$ is negative close to a Lifshitz solution.

The EPD theory \eqref{eq:action} admits the Lifshitz solutions (with $z>1$)
\begin{equation}
\label{eq:solution}
 ds^2  =  -\frac{1}{r^{2z}}dt^2+\frac{1}{r^2}\left(dr^2+dx^2+dy^2\right) \spa 
 B =  A_0\frac{1}{r^{z}}dt \spa 
 \Phi  =  \Phi_\star\,\;.
 \end{equation} 
Here, $\Phi_*$ is  constant, $ A_0^2 = 2(z-1)/(z Z_0)$ and we have the conditions
\be
 V_0  =  -(z^2+z+4)\sp \frac{W_0}{Z_0}  = 2z \spa V_1  =  (za+2b)(z-1) \,\;,
 \label{eq:conditionV1}
\ee
where  $a=Z_1/Z_0$, $b=W_1/W_0$ and $Z_i,W_i,V_i$ are the Taylor coefficients of the functions $Z, W, V $ around
$\Phi_*$, the value of which, together with $z$, is determined by the first two equations in \eqref{eq:conditionV1}. 
The third equation in \eqref{eq:conditionV1} is an extra condition that makes Lifshitz a non-generic solution of the family of actions \eqref{eq:action}. We note that there are also solutions of the EPD model with a running scalar whose metric is a Lifshitz space-time \cite{Gouteraux:2012yr,Gath:2012pg}, which will not be considered here.

To define our notion of  asymptotically locally Lifshitz space-times  it will prove convenient to write
\begin{equation}\label{eq:gaugemetric}
ds^2 = \frac{dr^2}{R(\Phi)r^2}-E^0E^0+\delta_{ab} E^a E^b\,,\qquad B_M = A_M-\partial_M\Xi\,\;,
\end{equation}
with the boundary at $r=0$. 
Our boundary conditions can then be summarized as%
\footnote{The recent article \cite{Chemissany:2014xsa} proposes what seems to be a different notion of AlLif space-times. 
We will comment on this difference in \cite{Hartong:2015wxa,Hartong:2014}.}
\be
\label{bc1}
E^0_\mu \simeq r^{-z}  \alpha_{(0)}^{1/3}\tau_\mu \spa E^a_\mu \simeq r^{-1} \alpha_{(0)}^{-1/3} e^a_\mu  \spa
A_\mu - \alpha (\Phi) E^0_\mu \simeq -r^{z-2} \tilde m_\mu \,\;,
\ee
\be
\label{bc2} 
\Xi \simeq -r^{z-2} \chi \spa A_r \simeq -(z-2) r^{z-3} \chi \spa \Phi \simeq r^\Delta \phi\,\;,
\ee
where $R (\Phi) \simeq R_{(0)}$ and $\alpha (\Phi) \simeq \alpha_{(0)}$ with $R_{(0)}$ and $\alpha_{(0)}$ functions of the boundary coordinates and we note for completeness that $A_\mu \simeq r^{-z} \alpha_{(0)}^{4/3} \tau_\mu$. Here the symbol $\simeq$ denotes
leading order in the near-boundary $r$-expansion.  
We will also need the inverse vielbeins
\be
E_0^\mu \simeq - r^z \alpha_{(0)}^{-1/3} v^\mu \spa E_a^\mu \simeq r \alpha_{(0)}^{1/3}e_a^\mu\,\;,
\ee
satisfying the orthogonality relations
\be
\label{eq:ortho}
v^\mu\tau_\mu=-1\,,\qquad v^\mu e_\mu^a=0\,,\qquad e^\mu_a\tau_\mu=0\,,\qquad e^\mu_a e_\mu^b=\delta^b_a\,\;.
\ee
As derived in detail in \cite{Hartong:2014}, it turns out that 
the equations of motion fix the form of $R_{(0)}$ and $\alpha_{(0)}$, so these are not independent sources. We now comment on the origin and motivation of the boundary conditions \eqref{bc1}, \eqref{bc2} as well as
the conditions coming from requiring a leading order solution of the equations of motion  of the model \eqref{eq:action}.

{\bf Dilaton}. 
First of all, in the condition for the dilaton $\Phi$ we allow for a weight  $\Delta\ge 0$.
We often encounter functions of $\Phi$ such as $Z$, $W$ and $V$. In order to solve the equations of motion near the boundary we need to expand these function around $\Phi=\Phi_\star$. These expansions depend on whether $\Delta>0$ or $\Delta=0$. By a shift in $\Phi$ we will take from now on the Lifshitz point to be at $\Phi_*=0$. The value of $\Delta$ can be computed by looking at radial perturbations around a pure Lifshitz solution. However as we will not need its explicit value we will not perform this analysis.

{\bf Metric}. 
Turning to the metric, we note that  we keep a non-trivial radial `lapse' function $R$, and hence we do in general not work in radial gauge which would mean $R=\text{cst}$ as is done for the AdS case. 
The near boundary ($r=0$) behavior is such that the powers in $r$ are not more divergent than for a pure Lifshitz solution. The need to work in a non-radial gauge, controlled by the function $R$, was noticed in \cite{Christensen:2013lma,Christensen:2013rfa} and is
reconfirmed in our more general model here.  
The form of $R_{(0)}$ is fixed by the near boundary behavior of the dilaton, i.e. whether $\Delta=0$ or $\Delta>0$, and the equations of motion. 
The fall-off conditions for the vielbeins are standard and the same as in e.g. \cite{Ross:2011gu} except that we will not impose by hand that $\tau_\mu$ is hypersurface orthogonal (HSO), and let the equations of motion
determine it.  

In fact the equations of motion show that 
for  $z>2$ the vielbein $\tau_\mu$ must be HSO, i.e. 
$\omega^2=0$ where 
 \begin{equation}\label{eq:twistHSOrelation}
\omega^2 =\frac{1}{2}\left(\varepsilon^{\mu \nu \rho }\tau_{\mu}\partial_\nu \tau_{\rho}\right)^2 \,,
\end{equation}
is the twist of $\tau_\mu$, where $\varepsilon^{\mu \nu \rho}$ is the boundary inverse Levi-Civit\`a tensor. 
In this case, the leading order equations of motion do not fix $R_{(0)}$ and $\alpha_{(0)}$. However, this can be accomplished for
 $1<z \le 2$ which is the case on which we focus.  The solution splits into four branches, 
i) $1<z<2$ and $\Delta>0$,  ii) $1<z<2$ and $\Delta=0$, iii) $z=2$ and $\Delta>0$ and iv) $z=2$ and $\Delta=0$
(details are given in \cite{Hartong:2014}). Here we note that in the first two cases there is no HSO constraint, in the third
case $\tau$ is HSO  and in the fourth case, there are two further possibilities depending on whether $W=4Z^{2/3}$ or not. In the former case we find that $\tau_\mu$ must be HSO, and in the latter case there is a constraint involving the source
$\phi$ 
\begin{equation}\label{eq:constraintrescaled}
\omega^2 =-2(Z(\phi))^{2/3}+\frac{1}{2}W(\phi)\,,
\end{equation}
This constraint parallels the constraint found in the $z=2$ model of \cite{Christensen:2013lma,Christensen:2013rfa},
which is closely related to the present model at $z=2$.

{\bf Vector field and St\"uckelberg scalar}.  For the pure Lifshitz solution the vector $B_\mu$ is proportional to $\tau_\mu$ as can be seen from \eqref{eq:solution}. We therefore  let $B_\mu \simeq r^{-z} \alpha_{(0)}^{4/3} \tau_\mu$ and since both $B_\mu$ and
$\alpha E^0_\mu$ have the same near-boundary behavior, we consider the linear combination $B_\mu - \alpha E^0_\mu$,
which has not been fixed so far. A relatively straightforward analysis \cite{Hartong:2015wxa,Hartong:2014} that uses bulk local Lorentz transformations
then fixes $B_\mu - \alpha E^0_\mu \simeq -r^{z-2}  M_\mu$, which is compatible with what is known about the  $z=2$ case discussed in \cite{Christensen:2013lma,Christensen:2013rfa}.  It is also interesting to note that, using the results of  e.g. \cite{Liu:2014tra},
this also works for $z=1$.  To address the near-boundary behavior of the radial component of $B_M$ we 
use the St\"uckelberg decomposition in \eqref{eq:gaugemetric},  invariance under the gauge transformations
$\delta A_M = \partial_M \Lambda$, $\delta \Xi = \Lambda$ and the decomposition 
$M_\mu = \tilde m_\mu - \partial_\mu \chi$.  The gauge choice for $A_r$ in \eqref{bc2} then follows if we expand
$\Lambda \simeq -r^{z-2} \sigma$.  The vector $\tilde m_\mu$ is what we 
call the boundary gauge field, observed for the first time in \cite{Christensen:2013lma,Christensen:2013rfa}. It plays a crucial role in the identification of the boundary geometry discussed below.

\section{Sources, torsional Newton-Cartan geometry and Schr\"odinger symmetry \label{sec:sources} }

We  now discuss  the transformation properties of the sources appearing in \eqref{bc1}, \eqref{bc2} that are induced
by local bulk symmetries. These consist of local tangent space transformations, gauge transformations and bulk diffeomorphisms.
By expanding bulk local Lorentz transformations near the boundary we see that because $z>1$ the timelike vielbein blows
up faster near the boundary than the spacelike ones, i.e. the local light cones flatten out. As a  result the Lorentz group contracts to the Galilei group so that $r \rightarrow 0$ is like sending the speed of light to infinity. Gauge transformations were already
discussed above and the relevant bulk diffeomorphisms are the  Penrose--Brown--Henneaux (PBH) transformations \cite{Penrose:1986ca,Brown:1986nw}, which preserve the form of the metric, i.e. the fact that $Rg_{MN}$ is in radial gauge. 
We then arrive at the following transformations of the boundary fields
\be
\label{eq:Schvartau}
\delta\tau_\mu  =  z\Lambda_D \tau_{\mu}  \spa 
\delta e^a_\mu  =  \lambda^a \tau_\mu + \lambda^{a}{}_b e^b_\mu + \Lambda_D e^a_\mu\,\;,
\ee
\be
\delta \tilde m_\mu =  \lambda^a e_a^\mu + \partial_\mu \sigma + (2-z) \Lambda_D \tilde m_\mu +(2-z) \chi \partial_\mu 
\Lambda_D \,\;,
\ee
\be
\label{eq:Schvarchi}
\delta \chi = \sigma +(2-z) \Lambda_D \chi  \spa \delta \phi =- \Delta \Lambda_D \phi\,\;,
\ee
\be
\label{eq:Schvarv}
\delta v^\mu = \lambda^a e_a^\mu - z \Lambda_D v^\mu
\spa \delta e^\mu_a = \lambda_a{}^b e^\mu_b - \Lambda_D e^\mu_a \,\;,
\ee
where for brevity we have omitted diffeomorphisms which act as Lie derivatives. 
Here $\lambda^a$  correspond to Galilean boosts ($G$), $\lambda_a{}^b$ to spatial rotations ($J$), $\Lambda_D$ to dilatations ($D$) and $\sigma$ to gauge transformations ($N$).  

Since we are working in a vielbein formalism when we consider variations of the on-shell action with respect to the boundary vielbeins we must decompose the boundary gauge field $ \tilde m_\mu = \tilde m_0 \tau_{\mu}+ \tilde m_a e_{\mu}^a$. Our sources are thus as summarized in table \ref{table:scalingdimensionssources} together with their scaling dimensions (dilatation weights).  This statement is modulo the possible $z=2$ constraints of HSO of $\tau_\mu$ or   \eqref{eq:constraintrescaled}. Note also that one either chooses the set $(\tau_\mu, e^a_\mu)$ or $(v^\mu, e_a^\mu)$.   It is instructive to count the sources taking into account the symmetries. We have in total 14 components (see table \ref{table:scalingdimensionssources} and omit $(v^\mu, e_a^\mu)$) and there are 8 local symmetry parameters contained in \eqref{eq:Schvartau}--\eqref{eq:Schvarv} and finally for $z=2$ we always have one constraint. This leaves us with $14-8=6$ free sources for $1<z<2$ and $5$ free sources for $z=2$. For the massive vector model, i.e. for $Z$, $W$ and $V$ constant and no $\Phi$, we count $5$ free sources for $1<z<2$ and $4$ when $z=2$.
The dual vevs and their scaling dimensions will be discussed further below.

\begin{table}[h!]
      \centering
      \begin{tabular}{|c|c|c|c|c|c|c|c|c|}
      \hline
source & $\phi$ & $\tau_\mu $ & $e_\mu^a $ & $v^\mu$ & $e_a^\mu$ & $ \tilde m_0 $ & $ \tilde m_a $  & $\chi $\\
  \hline
 scaling dimension &$\Delta$&$-z$&$-1$&$z$&$1$&$2z-2$&$z-1$&$z-2$    \\
         \hline
           \end{tabular}
      \caption{Sources and their scaling dimensions.} \label{table:scalingdimensionssources} 
\end{table}
{\bf Torsional Newton-Cartan geometry}. 
In the $z=2$ model of \cite{Christensen:2013lma,Christensen:2013rfa} it was observed that the boundary geometry is described by Newton--Cartan (NC) geometry extended with the inclusion of a specific torsion tensor and dubbed torsional Newton--Cartan (TNC) geometry. 
We now show that this is also the case in our general $z$ Lifshitz model.  To this end it will be very convenient to introduce
the following Galilean boost invariant objects
\be 
 \hat v^\mu  =   v^\mu-h^{\mu\nu}M_\nu  \spa \hat e_\mu^a  = e_\mu^a - M_\nu e^{\nu a} \tau_\mu   
   \spa \tilde\Phi  =  -v^\mu M_\mu+\frac{1}{2}h^{\mu\nu}M_\mu M_\nu\,\;,
 \ee
 \be 
  h^{\mu\nu}  =  \delta^{ab}e^\mu_a e^\nu_b \spa 
  \bar h_{\mu\nu} =  \delta_{ab}e^a_\mu e^b_\nu-\tau_\mu M_\nu-\tau_\nu M_\mu\,\;.
\ee

The vielbeins $\hat v^\mu$, $\hat e_\mu^a$, $\tau_\mu$, $e^\mu_a$ satisfy the same orthogonality relations as in \eqref{eq:ortho}. 
Note  in particular that $\tilde\Phi$ is the component of $M_\mu$ that cannot be removed by boost transformations. This is a new
source that appeared for the first time in  \cite{Christensen:2013lma,Christensen:2013rfa} and was previously not identified
in the Lifshitz literature. It is crucial to keep the full $M_\mu$ in the formalism to identify the boundary geometry and the full
set of symmetries in the on-shell action.  We refer to $\tilde\Phi$ as the Newtonian potential for reasons explained in \cite{Bergshoeff:2014uea}. 

Out of the quantities we have defined above we can build  an affine connection $\Gamma^{\rho}_{\mu \nu}$ that is invariant under the local symmetries $(G,J,N)$ and that satisfies metric compatibility with respect to the metric tensors $\tau_\mu$ and $h^{\mu \nu}$. This takes the simple form 
\be
\Gamma^{\rho}_{\mu\nu}=-\hat v^\rho\partial_\mu\tau_\nu+\frac{1}{2}h^{\rho\sigma}\left(\partial_\mu\bar h_{\nu\sigma}+\partial_\nu\bar h_{\mu\sigma}-\partial_\sigma\bar h_{\mu\nu}\right)\,\;,
\ee
so that the torsion tensor is given by 
\be
\label{eq:torsion}
\Gamma^\rho_{[\mu\nu]}=-\frac{1}{2}\hat v^\rho(\partial_\mu\tau_\nu-\partial_\nu\tau_\mu)\,\;.
\ee 
The connections for rotations $\Omega_\mu{}^a{}_b$ and boosts $\Omega_\mu{}^a$ are defined via the covariant derivatives and vielbein postulates. 
For example 
\be
\label{eq:VP}
\mathcal{D}_{\mu} e_{\nu}^{a} =  \partial_\mu e_{\nu}^{a}-\Gamma^{\rho}_{\mu \nu}e_{\rho}^{a}-\Omega_{\mu}{}^{a}\tau_{\nu}-\Omega_{\mu}{}^{a}{}_{b}e_{\nu}^{b} = 0\,\;.
\ee
The remaining three vielbein postulates have a similar form.  We also note that the covariant derivative acting on $M^a$, denoted by $D_\mu M^a$, is given by\footnote{We reserve the notation $\mathcal{D}_{\mu}M^{a}$ for a slightly different covariant derivative defined in \cite{Bergshoeff:2014uea}.}
\begin{equation}\label{eq:covderM}
D_{\mu}M^a = \partial_\mu M^a-\Omega_{\mu}{}^a-\Omega_\mu{}^a{}_b M^b\,.
\end{equation}
In \cite{Bergshoeff:2014uea} it is shown how to go further and make covariant derivatives with respect to local dilatations by introducing a new connection $b_\mu$, which leads to the existence of a local special conformal symmetry.

An important special case of TNC geometry is obtained by requiring $\tau_\mu$ to be HSO, which was called twistless torsional Newton-Cartan (TTNC) geometry in \cite{Christensen:2013rfa} since in that case the twist \eqref{eq:twistHSOrelation} vanishes. This does not necessarily imply that the torsion \eqref{eq:torsion} of the metric compatible connection is zero, but that there is zero torsion on spatial slices. This is the boundary geometry for $z>2$ and for many $z=2$ cases depending on the details of the model. In the case of TTNC geometry we can always apply a local dilatation to turn the geometry into a Newton--Cartan geometry for which $\tau$ is closed so that the torsion \eqref{eq:torsion} vanishes. Hence the TTNC torsion can be viewed as resulting from dilatation invariance.  
 
 {\bf Schr\"odinger symmetry}. We will next discuss the emergence of Schr\"odinger transformations acting on the sources. The transformations \eqref{eq:Schvartau}--\eqref{eq:Schvarchi} under the $G$, $J$, $N$, $D$ transformations can be compactly written as
\begin{equation}\label{eq:YMtrafo}
\delta{\mathcal{A}}_{\mu}=\partial_\mu\Sigma+[{\mathcal{A}}_{\mu}\,,\Sigma]\,,
\end{equation}
where $\mathcal{A}_{\mu}$ and $\Sigma$ are Schr\"odinger Lie algebra-valued and given by
\begin{eqnarray}
\mathcal{A}_\mu & = & H\tau_{\mu}+P_ae_{\mu}^a+G_a\omega_{\mu}{}^a+\frac{1}{2}J_{ab}\omega_{\mu}{}^{ab}+Nm_{\mu}+Db_{\mu}\,,\label{eq:Schconnection}\\
\Sigma & = & G_a\lambda^a+\frac{1}{2}J_{ab}\lambda^{ab}+N\sigma+D\Lambda_D\,,\label{eq:Sigma}
\end{eqnarray}
which involves the dilatation connection $b_\mu$ mentioned just below \eqref{eq:covderM}, with the Schr\"odinger algebra given by 
\begin{equation}\label{eq:Schalgebra}
\begin{array}{ll}
\left[D\,,H\right] = -zH\,,   &  \left[D\,,P_a\right] = -P_a\,,\\
\left[D\,,G_a\right] = (z-1)G_a\,,   &  \left[D\,,N\right] = (z-2)N\,,\\
\left[H\,,G_a\right] = P_a\,, & \left[P_a\,,G_b\right] = \delta_{ab}N\,,\\
\left[J_{ab}\,,P_c\right] = \delta_{ac}P_b-\delta_{bc}P_a\,, &
\left[J_{ab}\,,G_c\right] = \delta_{ac}G_b-\delta_{bc}G_a\,,\\
\left[J_{ab}\,,J_{cd}\right] = \delta_{ad}J_{bc}-\delta_{ad}J_{bc}-\delta_{bc}J_{ad}+\delta_{bd}J_{ac}\,. &
\end{array}
\end{equation}
with $\tilde m_\mu = m_\mu - (z-2) \chi b_\mu$ and $\chi$ transforming as in \eqref{eq:Schvarchi}.

In \cite{Bergshoeff:2014uea} it is shown how to include furthermore the local time and space translations generated by $H$ and $P_a$ in the expression for $\Sigma$ in such a way that \eqref{eq:YMtrafo} describes the diffeomorphisms generated by $\xi^\mu$. This is achieved via so-called curvature constraints whose solutions provide us with expressions for the connections $\omega_\mu{}^a{}_b$, $\omega_\mu{}^a$ and $e^\mu_a b_\mu$ in terms of $\tau_\mu$, $e_\mu^a$ and $M_\mu$ with $\omega_\mu{}^a{}_b$, $\omega_\mu{}^a$ dilatation covariant generalizations of $\Omega_\mu{}^a{}_b$, $\Omega_\mu{}^a$ defined earlier in \eqref{eq:VP}. The resulting technique is referred to as gauging the Schr\"odinger algebra which can be viewed as an extension of the work on gauging the Bargmann algebra \cite{Andringa:2010it} extended to include dilatations since the Bargmann algebra plus local dilatations gives the Schr\"odinger algebra.

Once we have imposed the curvature constraints an extra symmetry, the $K$ transformation, emerges which allows us to transform away the $\hat v^\mu b_{\mu}$ part of the $b_{\mu}$ connection (which was not fixed by the curvature constraints)\footnote{For $z=2$ this symmetry also exists before imposing the curvature constraints and amounts to working with the full $z=2$ Schr\"odinger Lie algebra, i.e. \eqref{eq:Schalgebra} with $z=2$ extended to include the special conformal generator $K$.}.  We refer the reader to \cite{Bergshoeff:2014uea} for details.

Hence, the entire boundary geometry including the transformations under diffeomorphisms can be obtained by gauging the entire local Schr\"odinger algebra (in the presence of the 
St\"uckelberg scalar $\chi$) with critical exponent $z$ and imposing what are known as curvature constraints that make local time and space translations equivalent to diffeomorphisms. 
From this perspective the gauge connection $m_\mu$ defined via $\tilde m_\mu = m_\mu - (z-2) \chi b_\mu$ is the gauge field of the mass generator of the Bargmann subalgebra which has dilatation weight $2-z$. Since $m_\mu$ and $\tilde m_\mu$ have the same dilatation weight this provides another argument for the $r^{2-z}$ fall-off of the linear combination in \eqref{bc1}.

\section{Vevs and covariant Ward-identities}

Finally we turn our attention to the vevs obtained by varying the (renormalized) on-shell action with respect to the sources. We think of the fall-off conditions \eqref{bc1}, \eqref{bc2} as Dirichlet boundary conditions in that we assume that there exists a local counterterm action on top of the usual Gibbons--Hawking (GH) boundary term that must be added to \eqref{eq:action} consisting of intrinsic terms, such that the on-shell action is finite and the variation with Dirichlet boundary conditions vanishes on-shell. One such counterterm action has been constructed in \cite{Christensen:2013rfa}, but more generally the construction of this requires a great deal of work. However, we will show that, provided it exists, many properties such as the definition of the vevs, their transformation properties under the Schr\"odinger group as well as their Ward identities can be derived without knowing the counterterm action explicitly. At the same time, the natural nature of the fall-off conditions, experience with previous models and the relation between sources and TNC geometry
 strongly suggests that large classes of Lagrangians \eqref{eq:action} admit a finite number of local counterterms. The only form of non-locality we will consider is the usual local scale anomaly term that is proportional to $\log r$. If our assumptions about the counterterm action are not obeyed the theory is either non-renormalizable or Dirichlet boundary conditions are not allowed and we are not interested in those cases here. 

Given these assumptions the variation of the total action takes the form  
$\delta S_{\rm ren}  =    -\int_{\partial\mathcal{M}}d^3x\, e {\cal{V}}  \delta {\cal{X}} $ (plus an anomaly term $-\mathcal{A} \delta r/r$),
where the bulk fields are collected in ${\cal{X}} = \{ E_0^\mu, E_a^\mu, \varphi, A_a,  \Xi , \Phi \} $ and
${\cal{V}} = \{ {\cal{S}}_\mu^0 , {\cal{S}}_\mu^a , {\cal T}_\varphi, {\cal T}^a , {\cal T}_\Xi, {\cal T}_\Phi \}$ and where $e$ is the determinant of the matrix $(\tau_\mu, e_\mu^a)$.  Here we have omitted the equations of motion
and defined $\varphi = E_0^\mu (A_\mu  - \alpha E_\mu^0) $. 
As a consequence we find the following expansions for ${\cal{V}}$ whose  leading terms are the vevs
\be
{\cal{S}}_\mu^0 \simeq  r^2\alpha_{(0)}^{2/3} S^0_\mu \spa 
{\cal{S}}_\mu^a \simeq r^{z+1}  S_\mu^a \spa
{\cal T}_\varphi \simeq r^{4-z}\alpha_{(0)}^{2/3} T^0 \spa {\cal{T}}^a \simeq r^3 T^a\,\;, 
\ee
\be
{\cal T}_\Xi \simeq r^4\alpha_{(0)}^{1/3}\langle O_\chi\rangle \spa 
{\cal T}_\Phi \simeq r^{z+2-\Delta}\alpha_{(0)}^{1/3}\langle O_\phi \rangle \spa 
\mathcal{A}  \simeq r^{z+2}\alpha_{(0)}^{1/3}\mathcal{A}_{(0)}\,\,,
\ee 
so that the variation of the on-shell action is 
\be
\delta S^{\text{os}}_{\text{ren}}  =  \int d^3x e\left[-S^0_\mu\delta v^\mu+S^a_\mu\delta e^\mu_a+T^0\delta\tilde m_0+T^a\delta\tilde m_a +\langle O_\chi\rangle\delta\chi+\langle \tilde O_\phi\rangle\delta\phi-\mathcal{A}_{(0)}\frac{\delta r}{r}\right]\,.
\label{eq:onshellvariation}
\ee
This exhibits a vev in front of every $\delta$(source) and we furthermore have defined 
\begin{equation}
\langle\tilde O_\phi\rangle=\langle O_\phi\rangle+ \delta_{\Delta,0} \left[\frac{1}{3}v^\mu\left(S^0_\mu-T^0\tilde m_\mu\right)+\frac{1}{3}e^\mu_a\left(S^a_\mu-T^a\tilde m_\mu\right)\right]  \frac{d \ln \alpha_{(0)}}{d\phi}\,\;.
\end{equation}

According to section \ref{sec:alif} the sources are unconstrained for $1<z<2$ so that the variations in \eqref{eq:onshellvariation} are free while for $z=2$ we always have a constraint. The variation of the on-shell action needs to be discussed separately for each of these three cases \cite{Hartong:2014}, which we now briefly discuss. In the case that $\tau_\mu$ is HSO it can be shown that since there is one less source the number of vevs is also reduced by one. In the case that we have the constraint \eqref{eq:constraintrescaled} it can be shown that we have $\langle\tilde O_\phi\rangle =0$, and since we know that $\phi$ is a function of $\omega^2$, which involves derivatives, we expect that a source for an irrelevant operator has been switched off as derivatives of sources appear at subleading orders. This feature has also been observed in the model discussed in \cite{Christensen:2013rfa}.

Using general properties of the quantities ${\cal{V}}$ appearing in the variation of the on-shell action, we can find
from the bulk symmetries, the complete local transformations of the vevs
\be
\delta S^{0}_{\mu}  =   T^{0} \partial_\mu \sigma + 2\Lambda_D S^{0}_{\mu} + \ldots
\spa 
\delta S^{a}_{\mu}  =  \lambda^{a}S^{0}_{\mu}+\lambda_{b}^{a} S^{b}_{\mu}+T^{a} \partial_\mu \sigma +(z+1)\Lambda_D S^{a}_{\mu}+\ldots\,,\\
\ee
\be 
\delta T^{0}  =  (4-z)\Lambda_D T^{0}+\ldots \spa 
\delta T^{a}  =  \lambda^{a}T^{0}+\lambda^{a}{}_{b}T^{b}+3\Lambda_D T^{a}+\ldots \,,
\ee
\be 
\delta\langle O_\chi \rangle  =  4\Lambda_D \langle O_\chi\rangle+\ldots \spa 
\delta\langle O_\phi \rangle  =   \delta_{\Delta\,,0} \frac{d \ln \alpha_{(0)}}{d \phi } \lambda_{a}T^{a}+(z+2-\Delta)\Lambda_D\langle O_\phi\rangle+\ldots\,,
\ee 
where the dots denote terms containing Lie derivatives along $\xi^\mu$ and possibly derivatives of $\Lambda_D$.
As was the case with the sources, the vevs transform under the Schr\"odinger group.

{\bf Boundary energy-momentum tensor and mass current}. 
We define the boundary energy-momentum tensor as the gauge invariant Hollands--Ishibashi--Marolf (HIM) boundary stress tensor \cite{Hollands:2005ya} that is invariant under $G$, $J$, $N$ transformations. By the HIM tensor we mean the tensor $-S^{0}_{\nu } v^\mu+S^{a}_{\nu}e_{a}^\mu$ which is invariant under tangent space transformations and obtained by varying the vielbeins. This object is however not invariant under local $N$ transformations and we therefore consider a gauge invariant extension $T^\mu{}_{\nu}$
which is provided by 
\begin{equation}\label{eq:bdrystresstensor}
T^\mu{}_{\nu}=-\left( S^{0}_{\nu}+T^{0} \partial_\nu \chi \right)v^\mu+
\left(  S^{a}_{\nu}+T^{a} \partial_\nu \chi \right) e_{a}^\mu \,\,.
\end{equation}
The scaling dimension of $T^\mu{}_\nu$ is $z+2$ and hence it is marginal. We note that the boundary energy-momentum tensor defined this way is a $(1,1)$ tensor and we remind the reader that we cannot raise and lower indices. 

The vielbein components of the energy-momentum tensor $T^\mu{}_\nu$ correspond to energy density ($T^\mu{}_{\nu} \tau_\mu v^\nu$), momentum flux ($T^\mu{}_\nu \tau_\mu e^\nu_a$), energy flux ($T^{\mu}{}_\nu e_\mu^a v^\nu$) and stress ($T^\mu{}_\nu e_\mu^a e_b^\nu$), respectively (see also \cite{Ross:2009ar}). They are presented in table \ref{table:scalingdimensions2} along with their scaling dimensions. In a non-relativistic theory mass and energy are no longer equivalent concepts. The mass density and mass flux are then provided by $T^0$ and $T^a$, respectively, which are the tangent space projections of the  current $T^\mu$ given by
\begin{equation}
T^\mu=-T^0 v^\mu+T^a e^\mu_a\,.
\end{equation}
These are also listed in table \ref{table:scalingdimensions2}.  We point out that even though the energy flux has scaling dimension $2z+1$ and would thus appear to be an irrelevant operator for $z>1$ this is not a problem since it is constructed entirely from the relevant operators that make up \eqref{eq:bdrystresstensor} contracted with (inverse) vielbeins, which are sources.

\begin{table}[h!]
      \centering
      \begin{tabular}{|c|c|c|c|c|c|}
      \hline
     $T^\mu{}_{\nu} \tau_\mu v^\nu$ & $T^\mu{}_\nu \tau_\mu e^\nu_a$ & 
     $T^{\mu}{}_\nu e_\mu^a v^\nu$  & $T^\mu{}_\nu e_\mu^a e_b^\nu $ & 
      $T^\mu \tau_\mu$ &  $T^\mu e_\mu^a $  \\
  \hline
$ z+2$ & 3& $2z+1$ & $z+2$ &$4-z$ &3    \\
         \hline
           \end{tabular}
      \caption{Scaling dimensions of tangent space components of the energy-momentum tensor and mass current.}\label{table:scalingdimensions2}
\end{table}

{\bf Ward identities}.  Since there are different classes of on-shell variations depending on whether $1<z<2$ or $z=2$, $\Delta>0$ or $\Delta=0$ and $W=4Z^{2/3}$ or $W\neq 4Z^{2/3}$ we need to consider the Ward identities for each case separately. These are obtained by demanding invariance of the 
variation of the on-shell action \eqref{eq:onshellvariation} with respect to the transformations \eqref{eq:Schvartau}--\eqref{eq:Schvarv}
 as well as under diffeomorphisms. These invariances are consequences of the fact that the bulk theory is invariant under diffeomorphisms, gauge and local Lorentz transformations. It turns out that 
 the final expressions for the Ward identities are the same in all three cases but their derivations are case dependent.
 
The Ward identities associated with local tangent space transformations (boosts and spatial rotations) are
\be
-\hat e_{\mu}^{a}T^\mu+\tau_\mu e^{\nu a}T^{\mu}{}_{\nu} = 0 \spa 
 \hat e_{\mu}^{a}e^{\nu b}T^{\mu}{}_\nu-\left(a\leftrightarrow b\right) = 0 \,.
\ee
We thus see that these reduce the number of components by 3, since the boost Ward identity relates 
the mass flux to the momentum flux and the one corresponding to rotations makes the spatial stress symmetric. 
The Ward identity for gauge transformations is
\be
e^{-1}\partial_\mu\left(eT^\mu\right)=\langle O_\chi\rangle\,\,,
\ee
while the one for dilatations takes the form 
\be
-z\hat v^\nu\tau_\mu T^\mu{}_{\nu}+\hat e_{\mu}^{a}e^{\nu}_a T^\mu{}_\nu+2(z-1)\tilde\Phi\tau_\mu T^\mu
= \mathcal{A}_{(0)}\,\,.
\ee
This exhibits the $z$-deformed trace and an extra term coming from the Newtonian potential. 
Finally, we have the Ward identity corresponding to diffeomorphisms
\be 
\nabla_{\mu}T^\mu{}_{\nu}+2\Gamma^{\rho}_{[\mu\rho]}T^\mu{}_{\nu}-2\Gamma^{\mu}_{[\nu\rho]}T^\rho{}_{\mu} -T^\mu\hat e_{\mu}^{a}D_{\nu}M_{a}+\tau_\mu T^{\mu}\partial_\nu\tilde\Phi = 0 \,\,.
\ee
It is interesting to note that the last term has the expected form of a force arising from the 
coupling of the mass current to the gradient of the Newtonian potential.

\section{Discussion}

We conclude by discussing some relevant open problems and extensions of our results. 

First of all, we note that we  have focussed our attention entirely on the leading order terms in the asymptotic expansion. By looking at linearized perturbations around the Lifshitz vacuum one can obtain an ansatz for the near boundary $r$ expansion for solving the full non-linear equations. It would be interesting to carry out this analysis to learn more about the case $z>2$ and to compute the counterterms. Regarding the latter the current leading order results are expected to be sufficient to fix the non-derivative counterterms. The subleading terms also control the expression for the anomaly density $\mathcal{A}_{(0)}$. From symmetry arguments we know that this must be a $G$, $J$, $N$ invariant scalar with dilatation weight $z+2$ (see also earlier work \cite{Baggio:2011ha,Griffin:2012qx,Chemissany:2012du,Christensen:2013rfa}). It would be interesting to use the Schr\"odinger symmetries to fix its general form as much as possible. The linearized perturbations around a Lifshitz vacuum lead to the same number of sources and vevs but they have a different fall-off behavior than what we mean by sources and vevs in the full non-linear case. It would be interesting to study the weak field limit of the asymptotic expansion including some of its subleading terms to see how this comes about. 

For future research it would be interesting to uncover the mechanism that makes the Lifshitz holographic setup used here such that the boundary theory exhibits Schr\"odinger symmetry. Is that only true for Einstein gravity coupled to a bulk vector field? For example what would happen%
\footnote{We thank Jan de Boer for interesting discussions on this point.}
in the context of Horava--Lifshitz gravity/Einstein-aether theories \cite{Horava:2009uw,Jacobson:2010mx}? 
The Schr\"odinger algebra has an infinite extension in the form of the Sch\"odinger--Virasoro algebra, it would be interesting to see if this plays a role in dual field theories to gravity on asymptotically 3D bulk Lifshitz space-times.

We also remark that we have assumed that the asymptotic geometry has no logarithmically running dilaton. However, it is known \cite{Gouteraux:2012yr,Gath:2012pg} that our model, the EPD action \eqref{eq:action}, admits solutions with another exponent (denoted by $\zeta$ in \cite{Gouteraux:2012yr} and by $\alpha$ in \cite{Gath:2012pg}) turned on that controls the logarithmic running. It would be interesting to extend our analysis to this case (see also \cite{Khveshchenko:2014nka,Karch:2014mba} in this context). In another direction, it would be interesting to add charge to our holographic Lifshitz setup. 

Finally for the purpose of applications of holography to CMT it would be interesting to study Lifshitz black branes (with and without nonzero mass density $T^0$) and to use ideas similar to those of the AdS fluid/gravity correspondence \cite{Bhattacharyya:2008jc} to uncover the hydrodynamics of the boundary field theory.

\section*{Acknowledgments}
\label{ACKNOWL}

We would like to thank Eric Bergshoeff, Geoffrey Comp\`ere, Simon Ross, Jan Rosseel and especially Jay Armas, Matthias Blau, Jan de Boer and Kristan Jensen for many valuable discussions. The work of JH is supported in part by the Danish National Research Foundation project ``Black holes and their role in quantum gravity''. 
The work of EK was supported in part by European Union's Seventh Framework Programme under grant agreements (FP7-REGPOT-2012-2013-1) no 316165,
PIF-GA-2011-300984, the EU program ``Thales'' MIS 375734, by the European Commission under the ERC Advanced Grant BSMOXFORD 228169 and was also co-financed by the European Union (European Social Fund, ESF) and Greek national funds through the Operational Program ``Education and Lifelong Learning'' of the National Strategic Reference Framework (NSRF) under ``Funding of proposals that have received a positive evaluation in the 3rd and 4th Call of ERC Grant Schemes''. The work of NO is supported in part by  Danish National Research Foundation project ``New horizons in particle and condensed matter physics from black holes''.  JH wishes to thank CERN for its hospitality and financial support.

\renewcommand{\theequation}{\thesection.\arabic{equation}}


\providecommand{\href}[2]{#2}\begingroup\raggedright\endgroup

\end{document}